\documentclass[preprint,aps,tightenlines]{revtex4}

\usepackage{graphicx}
\usepackage{color}
\usepackage{dcolumn}
\usepackage{bm}
\newcommand{\be}{\begin{equation}}
\newcommand{\ee}{\end{equation}}
\begin{document}

\title{Non-adiabatic generation of a pure spin current in a 1D quantum ring with
spin-orbit interaction}

\author{Marian Ni\c t\u a}
\affiliation{National Institute of Materials Physics, P.O. Box MG-7,
Bucharest-Magurele, Romania}

\author{D. C. Marinescu}
\affiliation{Department of Physics and Astronomy, Clemson University, Clemson,
South Carolina 29634, USA}

\author{Andrei Manolescu}
\affiliation{School of Science and Engineering, Reykjavik University, Menntavegur 1,
IS-101 Reykjavik, Iceland}

\author{Vidar Gudmundsson}
\affiliation{Science Institute, University of Iceland, Dunhaga 3, IS-107 Reykjavik,
Iceland}

\begin{abstract}
We demonstrate the theoretical possibility of obtaining a pure spin
current in a 1D ring with spin-orbit interaction by irradiation
with a non-adiabatic, two-component terahertz laser pulse, whose
spatial asymmetry is reflected by an internal dephasing angle $\phi$.
The stationary solutions of the equation of motion for the density
operator are obtained for a spin-orbit coupling linear in the electron
momentum (Rashba) and used to calculate the time-dependent charge and
spin currents.  We find that there are critical values of $\phi$ at
which the charge current disappears, while the spin current reaches a
maximum or a minimum value.
\end{abstract}

\maketitle

\section{Introduction}

Obtaining and controlling spin currents in solid structures has been
one of the most important goals of spintronics research. In the recent
past, the focus of this endeavor has been on the creative use of the
spin-orbit interaction (SOI) that appears in systems with broken inversion
symmetry, be that via confinement (Rashba)\cite{rashba} or in the bulk
(Dresselhaus)\cite{dresselhaus}. The well known properties of quasi-one
dimensional rings to support persistent charge currents \cite{buttiker,
wendler, chakraborty} made them particularly appealing for spin-dependent
exploration. The underlying physical phenomenon leading to the creation of
persistent currents is the imbalance in the left/right charge carrying
states realized, initially, in the presence of a magnetic flux threaded
through the center of the ring. In a similar experimental setup, the
existence of persistent spin-currents was obtained in the presence of SOI
\cite{splettstoesser, souma, sheng}, when electron-electron interaction
are neglected. Both analytical \cite{splettstoesser} and numerical
results \cite{sheng}, point out that in the presence of a magnetic flux,
the charge and spin currents in the rings are simultaneously present.

In this paper, we revisit this problem from the perspective of creating
persistent charge and spin current by non-adiabatic methods. As before,
the original exploration of such ideas was focused on the generation
of charge currents various mechanisms able to realize of an imbalance
between left\/right momentum-carrying states\cite{moskalets1, moskalets2,
gudmundsson}. In particular, we are concerned with the application
of a ultrashort, terahertz frequency laser pulse, endowed with a
spatial asymmetry expressed through an internal dephasing angle. The
efficiency of this method for creating persistent charge currents has
been explored in Refs.~\onlinecite{gudmundsson,siga}. Its extension
to a ring endowed with SOI is discussed below. As we demonstrate, the
interplay between the spin orbit coupling that rotates the electron-spin
around the ring and the spatial asymmetry of the initial excitation
generates {favorable} conditions where the charge current disappears,
while the spin current reaches a maximum level. For both a dipolar and a
quadrupolar perturbation, we can find the critical values of the dephasing
angle for which this situation occurs.

\section{The ring model}

We consider a one-dimensional (1D) quantum ring of radius $r_0$
containing few electrons, endowed with a Rashba interaction, linear in
the electron momentum. In a discrete (tight-binding) representation the ring is reduced
to $N$ sites (points) distributed on a circle, whose angular coordinate
is given by $\theta_n=2n\pi/N$ with $n=1,...,N$ the site index.
The Hamiltonian describing the noninteracting electrons is written
in terms of the creation and annihilation operators $c^{\dagger}_{n\sigma}$
and $c^{}_{n,\sigma}$ associated with the single-particle states
$|n\sigma \rangle$, where $\sigma=\pm1$ is the spin index.  This Hamiltonian 
has been extensively discussed in literature 
\cite{{meijer},{splettstoesser},{souma},{sheng}}, so here we will write it directly:
\begin{eqnarray}\label{eq:h0}
H &=& V\left\{2\sum_{n,\sigma}c^{\dagger}_{n\sigma}c^{}_{n\sigma}-\sum_{n,\sigma}
\left[c^{\dagger}_{n\sigma}c^{}_{n+1\sigma}+ c^{\dagger}_{n\sigma}c^{}_{n-1\sigma}\right]\right\}  \nonumber\\
&-& iV_{\alpha}\sum_{n,\sigma,\sigma'}
\left[{\sigma}_r(\theta_{n,n+1})\right]_{\sigma \sigma'}
      c^{\dagger}_{n\sigma} c^{}_{n+1\sigma'}+{\rm h.c.} \,
\end{eqnarray}

The two energy scales of the problem are set by the hopping matrix
element $V=\hbar^2/2m^*a^2$, where $m^*$ is the effective
electron mass and $a=2\pi r_0/N$ is the discretization constant,
and by the Rashba coupling, of strength $\alpha$, which generates
$V_{\alpha}=\alpha/2a$.  The spin operator $\sigma_r(\theta)$ introduced in
Eq.~(\ref{eq:h0}) represents the local orientation of the electron spin
along the radius of the ring and is given by a linear combination of
the Pauli operators $\sigma_x, \sigma_y$, written for the azimuthal
coordinate $\theta_{n,n+1}=(\theta_n +\theta_{n+1})/2$,
\be
\sigma_r (\theta_{n,n+1})
= \sigma_x\cos\theta_{n,n+1} + \sigma_y\sin\theta_{n,n+1} \;.\label{eq:sigma_r}
\ee
\begin{figure}[htp]
\includegraphics[width=9cm]{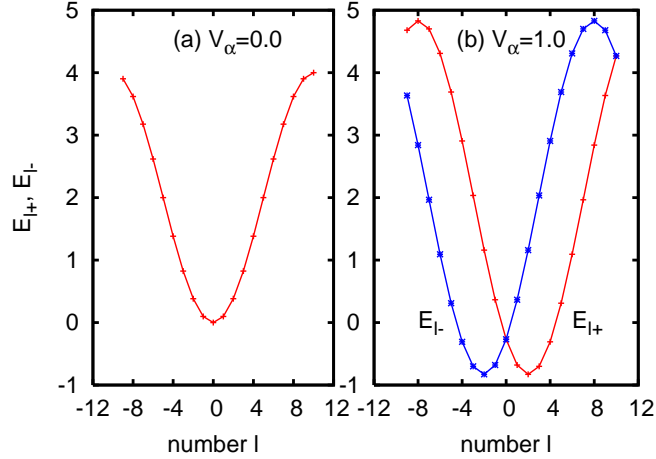}
\caption{(Color online) The eigenvalues $E_{l\pm}$ of the 1D quantum
ring with N=20 points vs. quantum number $l$. (a) $V_\alpha=0.0$, (b)
$V_\alpha=1.0$. The energy unit is $V=\hbar^2/2m^*a^2$.}
\label{Espec}
\end{figure}
The energy spectrum of the Hamiltonian, calculated for an even number
of sites, $N = 20$, is shown in Fig.\ \ref{Espec} for $V_\alpha=0$
and for $V_\alpha=1.0$ units of $V$. The split realized by SOI is
described by two eigenvalues, $E_{l+}$ (right arm) and $E_{l-}$ (left arm),
with $l=0, \pm 1, \pm 2, \cdots, \pm (N/2-1), N/2$,
\begin{eqnarray}\label{eigen1}
E_{l\pm}=\frac{E_l+E_{l\pm 1}}{2} +
\frac{E_l- E_{l\pm 1}}{2}\sqrt{1+\tan^2 2\theta_\alpha} \, ,
\end{eqnarray}
where $E_l=2V-2V\cos(2\pi l/N) \,$ is the degenerate eigenvalue in the absence of SOI and
$\theta_\alpha$ is given by
\begin{equation}
\tan 2\theta_\alpha=\frac{V_\alpha}{V\sin(\Delta\theta/2)} \,\, .
\end{equation}
The corresponding eigenvectors of the Hamiltonian (\ref{eq:h0}) are:
\begin{eqnarray}\label{eigenvectorsR}
&&|\Psi_{l+}\rangle=\frac{1}{\sqrt{N}}\sum_n e^{il\theta_n}
\left( \begin{array}{c}
\cos \theta_\alpha\\
- e^{i\theta_n}\sin \theta_\alpha
\end{array}
\right) |n\rangle \, , \nonumber \\
&&|\Psi_{l-}\rangle=\frac{1}{\sqrt{N}}\sum_n e^{il\theta_n}
\left( \begin{array}{c}
e^{-i\theta_n}\sin \theta_\alpha  \\
\cos \theta_\alpha
\end{array}
\right) |n\rangle \, .
\end{eqnarray}

The velocity operator is given by 
\begin{eqnarray}\label{vt}
v_\theta &=& r\dot{\theta}=r\frac{i}{\hbar}[H,\theta] \nonumber\\
&=& -\frac{Va}{\hbar} \Big\{
i\sum_{n,\sigma}c^{\dagger}_{n \sigma}c^{}_{n+1\sigma} 
+ \frac{V_{\alpha}}{V}\sum_{n,\sigma,\sigma'}
\left[{\bf \sigma}_r(\theta_{n,n+1})\right]_{\sigma \sigma'}
c^{\dagger}_{n \sigma} c^{}_{n+1 \sigma'}+ {\rm h.c.}\Big\} \, ,
\end{eqnarray}
and its eigenvectors are also $|\Psi_{l\pm}\rangle$, with the associated eigenvalues
\begin{eqnarray}\label{eigen2}
v_{l\pm}=\frac{v_l+v_{l\pm 1}}{2} +
\frac{v_l- v_{l\pm 1}}{2}\sqrt{1+\tan^2 2\theta_\alpha} \, ,
\end{eqnarray} 
where  $v_l=2 \left(Va/\hbar\right) \sin(2\pi l/N) \,$ is the velocity 
in the absence of SOI.

It is customary to introduce the tilt-spin operator $S_{2\theta_\alpha}$,
\begin{eqnarray}\label{spinteta}
S_{2\theta_\alpha}=\cos 2\theta_\alpha S_z -\sin 2\theta_\alpha S_r,
\end{eqnarray}
which is a linear combination of $S_z$ and $S_r$, the spin
operators for $z$ and radial directions, respectively.  The spinors
$|\Psi_{l\pm}\rangle$ are also eigenvectors of $S_{2\theta_\alpha}$
associated with eigenvalues $\pm \hbar/2$.

\begin{figure}
\includegraphics[width=8cm]{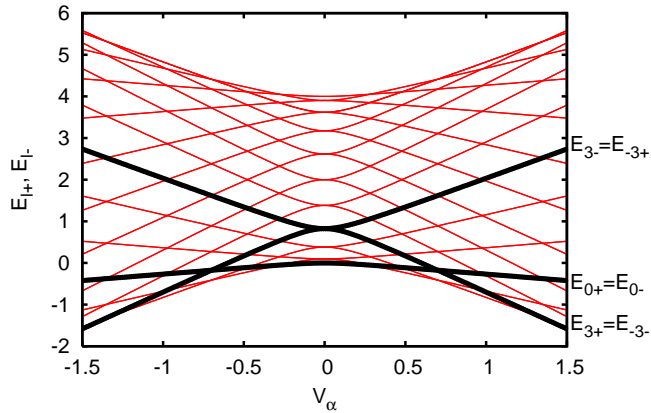}
\caption{Spectrum of the 1D quantum ring with N=20 points vs. Rashba
parameter $V_\alpha$.  The eigenvalues $E_{3\pm}$
and $E_0\pm$ are dotted to illustrate the lifting of degeneracy when $V_\alpha \neq 0$.}
\label{figura2}
\end{figure}

In the absence of SOI, all quantum states are four-fold degenerate, except
those at the edges of the spectrum which are only twice degenerate.
This can be seen in Fig.\ \ref{Espec}(a) where for $V_\alpha = 0$ the two
branches $E_{l+}$ and $E_{l-}$ coincide.  In this case the spin operator
$S_{2\theta_\alpha}$ becomes $S_z$ and $|\Psi_{l\pm}\rangle$ become
the  $\mid \uparrow\rangle$ and $\mid \downarrow\rangle$ eigenstates
of $S_z$.  For $V_\alpha \neq 0$, the energy spectrum becomes
broader and all states are twice
degenerate since $E_{l\pm}=E_{-l\mp}$. States at the crossing points
in the spectrum shown in Fig.\ \ref{figura2} preserve the four-fold
degeneracy.  Since these degenerate states carry opposite currents
$v_{l\pm}=-v_{-l\mp}$, the system can support a nonzero spin current.

\section{The time evolution}

At $t=0$ the quantum ring described above is exposed to a short terahertz
two-component pulse,
\begin{equation}\label{htn}
H_n(t)=A e^{-\Gamma t} \left[\sin (\omega_1t)  \cos \theta +\sin (\omega_2 t)  \cos n(\theta+\phi)\right],
\end{equation}
of duration $\sim\Gamma^{-1}$ and amplitude $A$ \cite{siga}. $n =
1,2$ describes the multipole order of the second component, while the
dephasing angle $\phi$ between the two components makes the external
perturbation asymmetric.  The terahertz scale of the excitation
frequencies $\omega_{1,2}$ is at least an order of magnitude larger than
the spin relaxation rates in InAs semiconductor heterostructures which
reaches values from anywhere between tens and hundreds of picoseconds
\cite{murdin, hall}.

The time evolution of the system's observables is determined by using
the density operator $\rho(t)$, which at $t>0$ satisfies a quantum
Liouville equation
\begin{equation}
i\hbar \rho(t) = \left[ H + H_n(t), \ \rho(t) \right ] \, .\label{eq:L}
\end{equation}
The initial condition is such that $\rho(t=0)$ represents the ground-state density
operator which is constructed in terms of the eigenvectors
$\mid\Psi_{l\sigma}\rangle$ shown in Eq.\ (\ref{eigenvectorsR}),
of the initial, time-independent Hamiltonian, Eq.~(\ref{eq:h0}):
\be\label{rho0}
    \rho(t=0)=\sum_{l\sigma} p_{l\sigma}
         \mid \Psi_{l\sigma}\rangle \langle \Psi_{l\sigma} \mid \, ,
\ee
where $p_{l\sigma}$ are the populations of the degenerate single-particle
states $\left( \sum_{l\sigma} p_{l\sigma}=1 \right)$.

For any $t>0$, Eq.~(\ref{eq:L}) is solved numerically and
$\rho(t)$ is obtained by using the Crank-Nicolson finite difference
method \cite{gudmundsson} with small time steps $\delta t\ll
\Gamma^{-1}f$.  The expectation value of any observable $O$ is then
calculated as $\langle O\rangle= {\rm Tr} \left(\rho(t)O\right)$.

\section{Results}

We define the charge current as $I^c=ev_\theta$ and the spin current
along a direction $\nu$ as $I^s_{\nu}=\frac{\hbar}{2} \left( {\bf
\sigma}_\nu v_\theta +v_\theta {\bf \sigma}_\nu \right)$.  Since the
spin operator $S_{2\theta_\alpha}$(\ref{spinteta}) commutes with the
unperturbed Hamiltonian (\ref{eq:h0}), we calculate the expectation
value of $I^s_{2\theta_\alpha}$ which corresponds to the direction
of the spin ${\bf e}_{2\theta_\alpha}=\cos 2\theta_\alpha {\bf e}_z
-\sin 2\theta_\alpha {\bf e}_r$.  To simplify the notation, we denote
$I^s(t)=\langle I^s_{2\theta_\alpha} \rangle$, $v_\theta(t)=\langle
v_\theta \rangle$ and $I^c(t)=ev_\theta(t)$.  Since $v_\theta$ and
$I^s_{2\theta_\alpha}$ commute with $H$, the charge and spin currents
become constant after the external perturbation vanishes.

To illustrate our results, we consider an InAs (electron effective
mass is $m^*=0.023m_e$)  quantum ring of radius $r_0=14$ nm. We
choose our Rashba parameter $V_\alpha=0.05$, which corresponds
to a SOI strength $\alpha=37.56$ meVnm, within the range of
experimentally determined values \cite{ganichev}.  The number of
sites used in the discretization is $N=20$, leading to a length
unit $a=4.4$ nm and an energy unit $V=85.6$ meV.  For a system
with $n_e=6$ electrons, in the many-particle, non-interacting
ground-state, the occupied states are $\Psi_{0\pm}$, $\Psi_{1\pm}$
and $\Psi_{-1\pm}$, with weights $p_{0\pm}=p_{1\pm}=p_{-1\pm}=1/n_e$
(and all the other $p_{l\sigma}=0$). The many-particle ground-state is
therefore nondegenerate. In this configuration, i.e. at $t=0$, the average
velocity is $v_\theta(t=0)=0$ and the spin current $I^s(t=0)=2\hbar
(v_{0+}+v_{1+}+v_{-1+})=-0.089Va/\hbar$.

\begin{figure}
\includegraphics[width=8cm]{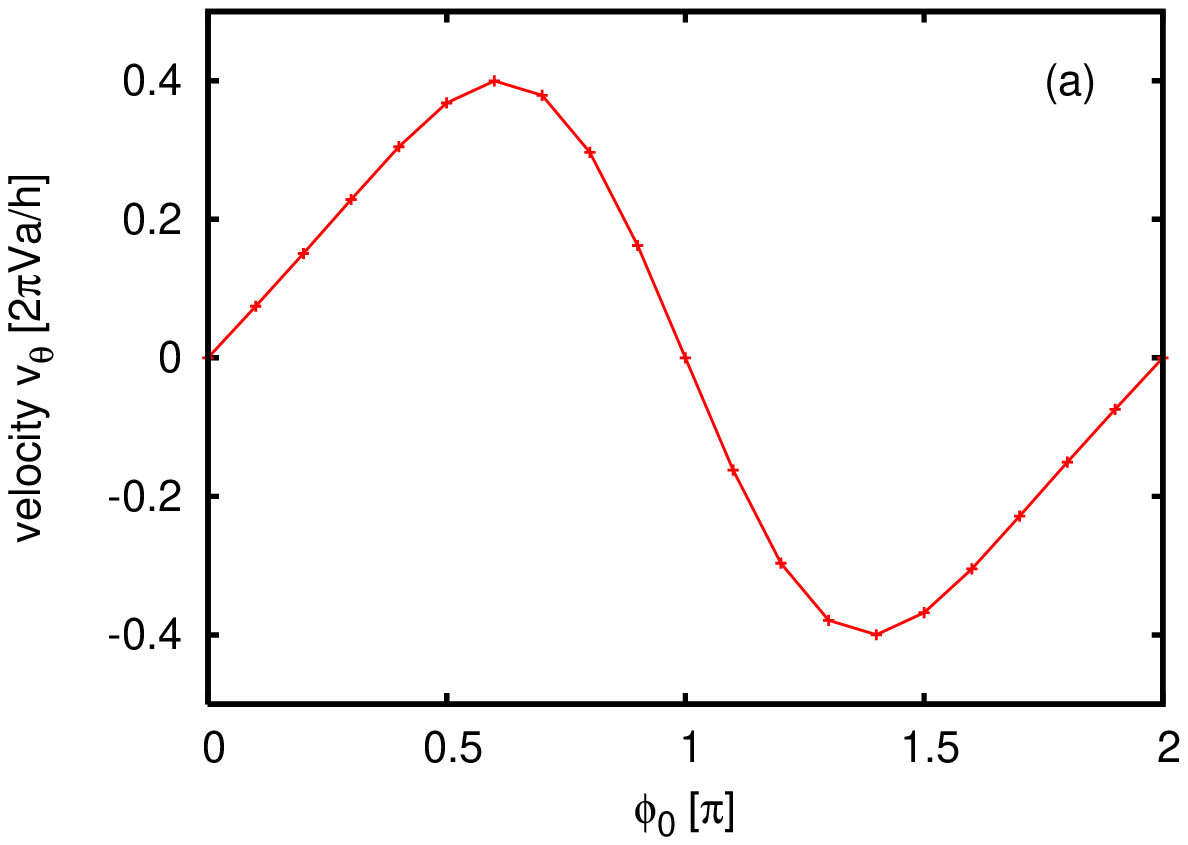}
\includegraphics[width=8cm]{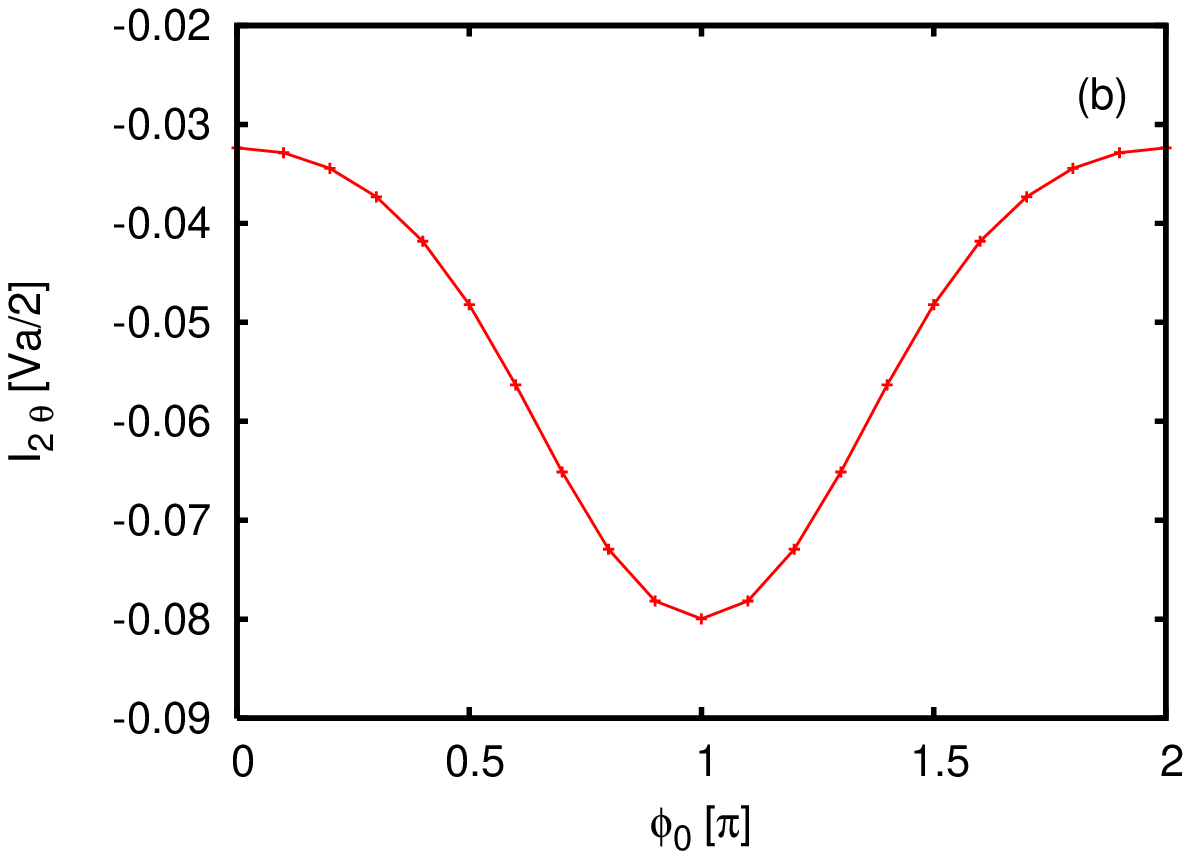}
\caption{The stationary values of the velocity $v_\theta$ and spin current
$I^s_{2\theta}$ after the external pulse vanishes (at $t\gg t_f$) vs.
the dephasing angle $\phi$.  The external pulse is $H_{1}(t)$, Eq.~(\ref{htn}) 
with $n=1$.}
\label{figura4}
\end{figure}

The first {external pulse} we consider is the superposition of two dipoles,
corresponding to $n = 1$ in Eq.~(\ref{htn}).  For the selected parameters
of the ring, in the absence of the SOI, we obtain the Bohr frequencies
$\hbar\omega_{0,1}=2.89$ meV and $\hbar\omega_{1,2}=8.60$ meV 
(defined as $\omega_{l,l'}=|E_l-E_{l'}|/\hbar$).
\begin{figure}
\includegraphics[width=8cm]{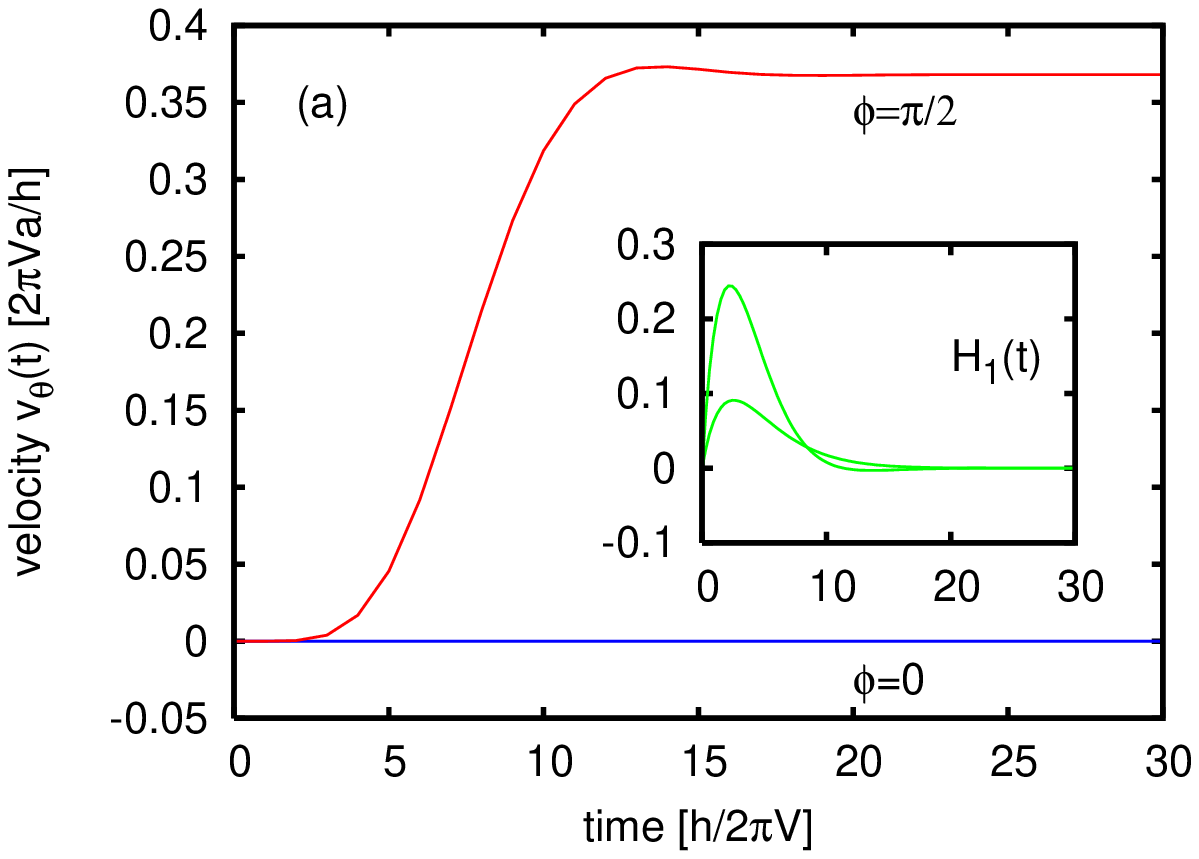}
\includegraphics[width=8cm]{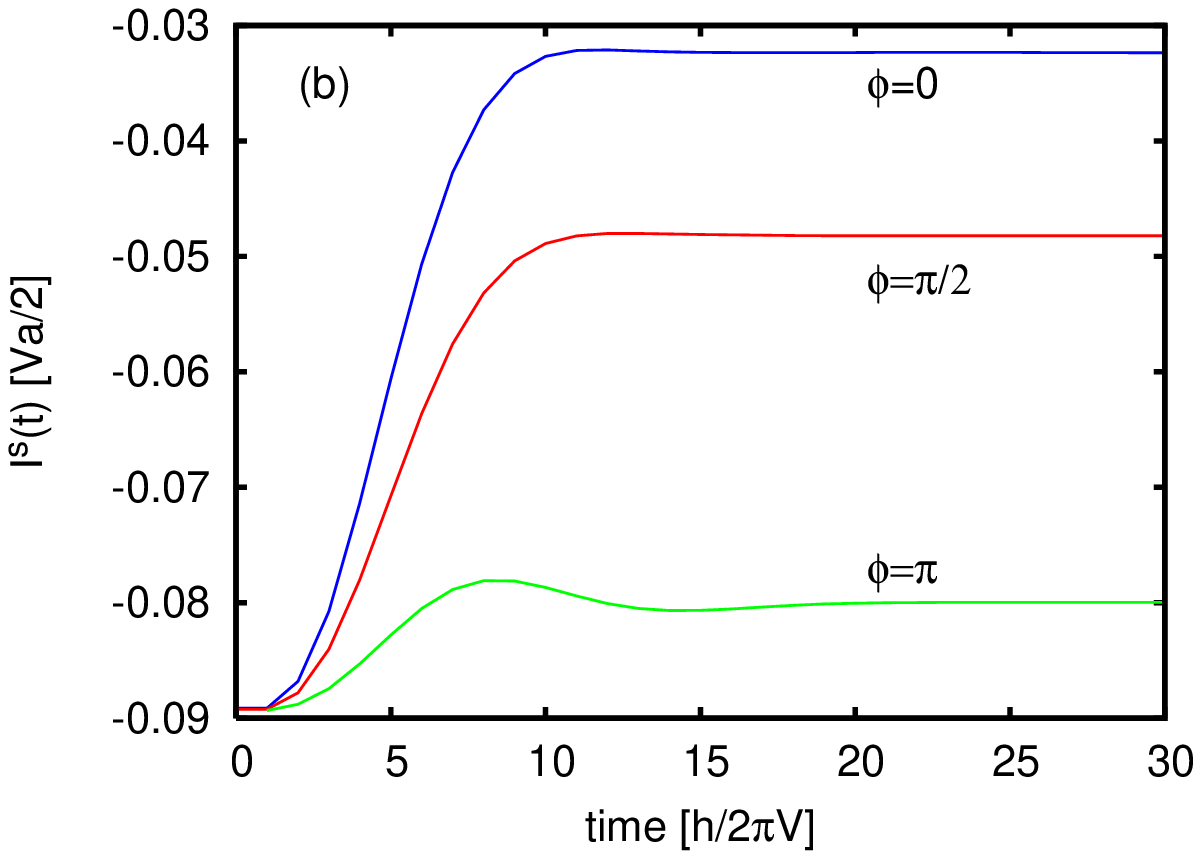}
\caption{(a) The time evolution of the azimuthal velocity
$v(t)=\langle v_\theta\rangle=I^c(t)/e$, (b) the spin current
$I^s(t)=\langle I^s_{2\theta_\alpha} \rangle$, for the selected dephasing
angles $\phi=0,\pi/2,\pi$.  The inset shows the two components of the
radiation pulse described by Eq.\ (\ref{htn}) with $n=1$, for $\theta=\phi=0$. The
lower peak corresponds to the first term and the higher peak to the second term.
}
\label{figura3}
\end{figure}
In Figs.~\ref{figura3} and \ref{figura4} we show the numerical results
obtained for frequencies $\hbar\omega_1=2.83$ meV, $\hbar\omega_2=8.11$
meV, with the attenuation factor $\Gamma=4\omega_1$, and amplitude
$A=67.68$ meV.  The duration of the pulse is $t_f\approx 0.5$ ps. Since the
pulse produces many-particle excited states, in our calculations $\omega_1$ and
$\omega_2$ are chosen to be slightly different from the Bohr frequencies, in
an effort to create a more realistic algorithm that reproduces closely
what happens in an experimental situation. Moreover, for $\omega_1$ and
$\omega_2$ exactly equal to the Bohr frequencies, the outcome is preserved.

The time evolution of velocity $v_\theta(t)$, proportional with
the charge current $I^c(t)$, and that of the spin current $I^s(t)$
are illustrated in Fig.~\ref{figura3} for dephasing angles $\phi=0$,
$\pi/2$ and $\pi$.  In Fig.~\ref{figura3}(a), we recover the result of
Ref.~\onlinecite{gudmundsson} where it was demonstrated that a charge
current can be non-adiabatically generated through the application
of a spatially asymmetric terahertz excitation. Thus,  for $\phi =
0$, $v_\theta = 0$, while for $\phi  = \pi/2$, $v_\theta \approx 0.37
V a/\hbar$.

In Figure \ref{figura3}(b) we present the time evolution of the spin
current, $I^s(t)$, corresponding to the spin projected on the proper
axis ${\bf e}_{2\theta_\alpha}$.  As previously stated, in the presence
of SOI, the initial state of the system has a nonzero spin current,
$I^s(0)\ne 0$.  On account of the external pulse, the electrons in the
ring are non-adiabatically excited, and the spin current evolves towards a
new steady-state value, $I_{2\theta}=I^s(t\gg t_f)$.  After the external
pulse vanishes, the spin current  $I_s(t)$ is constant in time, but its
amplitude varies with the dephasing angle.

We call $v_\theta$ and $I_{2\theta}$ the constant values of the velocity
and spin current after the perturbation vanished {(i.e. $v_{\theta}(t\gg
t_f)$ and $I^s(t\gg t_f)$ respectively).} They are plotted as a function
of the dephasing angle $\phi \in [0,2\pi]$ in Fig.~\ref{figura4}, but we
note that they reflect the periodicity in $\phi$ of the applied pulse.
For {$\phi_{v1}=0.6\pi$ and $\phi_{v2}=1.4\pi$ the charge current
is maximum and minimum, respectively.}  For $\phi$ equal to $0$ and
$\pi$ {no charge current flows through the ring ($v_\theta=0$) whereas
a spin current is present, $I_{2\theta}\neq 0$, being maximum for
$\phi_{s1}=0$ and minimum for $\phi_{s2}=\pi$.} These are the critical
{dephasing angles} for which only pure spin currents exist.  In this
case, the numerical results show that the states with opposite velocity,
$\Psi_{l+}$ and $\Psi_{(-l)-}$, are equally excited by the external pulse
netting zero {total} velocity and no charge current in the steady state.
This is not the case for intermediate angles where  states with opposite
velocity are asymmetrically excited and consequently, generating nonzero
values for {both} charge and spin currents.

\begin{figure}
\includegraphics[width=8cm]{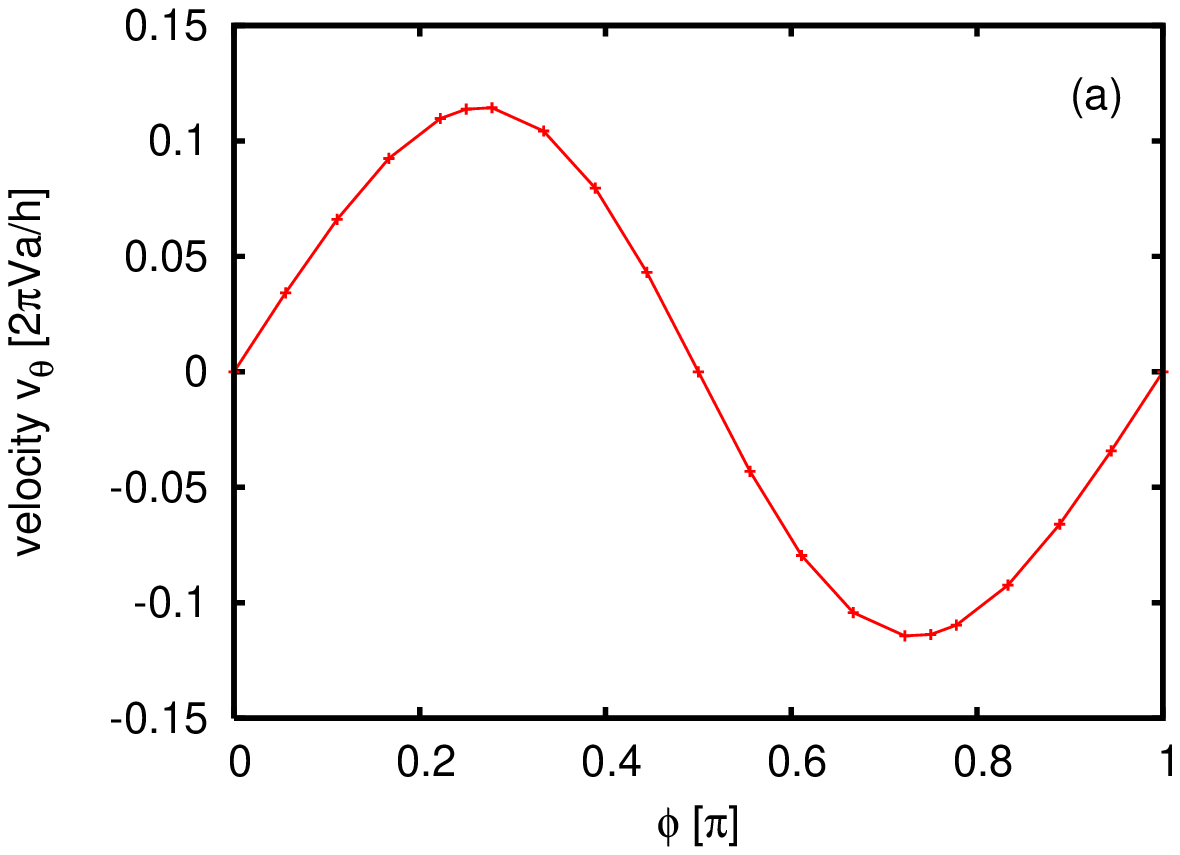}
\includegraphics[width=8cm]{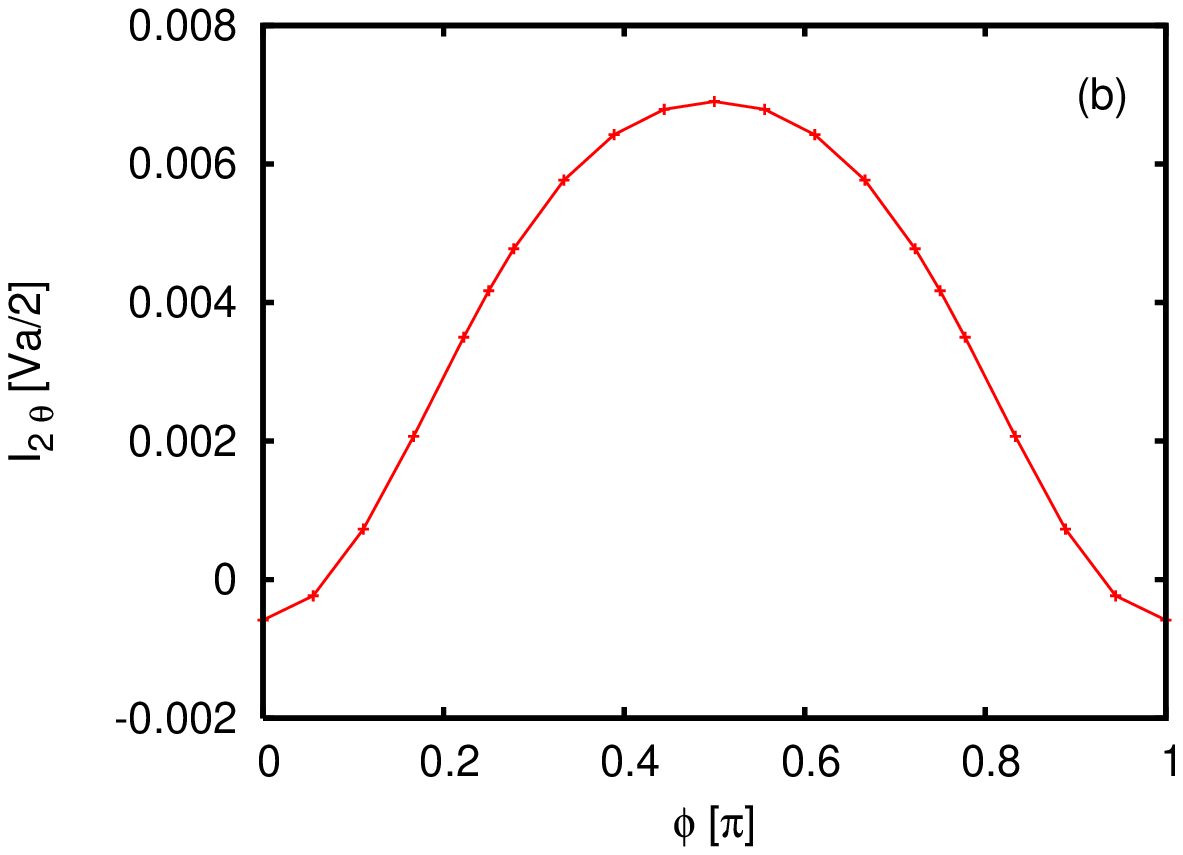}
\caption{The stationary values of the velocity $v_\theta$ and spin current
$I^s_{2\theta}$ after the external pulse vanishes (at $t\gg t_f$) vs.
the dephasing angle $\phi$.  The external pulse is $H_2(t)$, Eq.~(\ref{htn}) with
$n=2$.}
\label{figura5}
\end{figure}

In Fig.~\ref{figura5} we display $v_\theta$ and $I_{2\theta}$ obtained
by exciting the system with a pulse given by Eq.~(\ref{htn}) written for
$n = 2$, i.e. by a combination of a dipole and phase shifted quadrupole.
The frequencies $\omega_1$, $\omega_2$, the attenuation factor $\Gamma$,
and the amplitude $A$ remain the same as before.  Due to the quadrupolar
component of the pulse, the period of $v_\theta$ and $I_{2\theta}$
is halved to $\Delta\phi=\pi$.  The maximum and minimum values of
$v_\theta$ are reached {for $\phi_{v1}=0.28\pi$ and $\phi_{v2}=0.72\pi$
(nearly half of the previous values).
For the critical angles $\phi_{s1}=0$ and $\phi_{s2}=\pi/2$} no {charge}
current is induced in the ring ($v_\theta=0$), but the induced spin
current {reaches} extreme values, minima or maxima respectively.  Again,
in these situations only a pure spin current is induced in the ring.

\section{Conclusions}

In conclusion, we studied the non-adiabatic  {excitation} of spin
and charge currents in a 1D quantum ring in the presence of Rashba SOI.
The ring was {subjected to} an external pulse that is spatially
asymmetric, having two components with a relative dephasing
angle. We investigated {two models}, a dipole plus a rotated dipole
and a dipole plus a rotated quadrupole.  By numerical calculation, we
showed that for {certain} values of $\phi$ called $\phi_{v1}$
and $\phi_{v2}$ the induced charge current reaches {extreme}
values with $I^c(\phi_{v1})=-I^c(\phi_{v2})$.  Due to the presence of SOI
a nonzero spin current is also induced in the ring,  for both
pulse models, with amplitudes depending on the parameters of the pulse.
We found simple critical values of the dephasing angle, $\phi_{c1}$
and $\phi_{c2}$, for which the induced charge current disappears,
{whereas the spin current reaches maxima or minima.}
{The method may be used in practice to convert an optical signal
of variable amplitude
into a dissipationless (persistent) spin current for information transfer
purposes.  Similar results can be obtained with the Dresselhaus SOI instead
of the Rashba SOI, due to the equivalence of the corresponding Hamiltonians.}

\begin{acknowledgments}

This work was supported by the Icelandic Research Fund,
DOE  grant number DE-FG02-04ER46139,
Romanian PNCDI2 program Grant No. 515/2009 and Grant No. 45N/2009.
We acknowledge helpful discussions with Alexandru Aldea and Mugurel
\c{T}olea. M.N. is thankful to Clemson University, Reykjavik University,
and Science Institute - University of Iceland, for hospitality.

\end{acknowledgments}



\end{document}